\documentclass{article}

\usepackage{arxiv}

\usepackage[utf8]{inputenc} 
\usepackage[T1]{fontenc}    
\usepackage{hyperref}       
\usepackage{url}            
\usepackage{booktabs}       
\usepackage{amsfonts}       
\usepackage{nicefrac}       
\usepackage{microtype}      
\usepackage{lipsum}		
\usepackage{graphicx}
\usepackage[round]{natbib}
\usepackage{doi}

\usepackage{bm}
\usepackage{color}
\usepackage{subcaption}
\usepackage{bm}
\usepackage{amsmath}
\usepackage{amssymb}
\usepackage{algorithm,algorithmic}

\newcommand{\argmin}{\mathop{\rm arg~min}\limits}
\newcommand{\indep}{\perp \!\!\! \perp}

\title{Rules Ensemble Method with Group Lasso for Heterogeneous Treatment Effect Estimation}

\author{
    Ke Wan \\
	Department of Medical Data Science\\
	Wakayama Medical University\\
    \texttt{wane19911017@gmail.com} \\
    \And
	Kensuke Tanioka \\
	Department of Biomedical Sciences and Informatics\\
	Doshisha University\\
	\texttt{ktanioka@mail.doshisha.ac.jp} \\
	\And
	Toshio Shimokawa \\
	Department of Medical Data Science\\
	Wakayama Medical University\\
	\texttt{toshibow2000@gmail.com} \\
}

\renewcommand{\shorttitle}{}

\begin{document}
\maketitle

\begin{abstract}
The increasing scientific attention given to precision medicine based on real-world data has led many recent studies to clarify the relationships between treatment effects and patient characteristics. However, this is challenging because of ubiquitous heterogeneity in the treatment effect for individuals and the real-world data on their backgrounds being complex and noisy. Because of their flexibility, various heterogeneous treatment effect (HTE) machine learning (ML)estimation methods have been proposed. However, most ML methods incorporate black-box models that hamper direct interpretation of the interrelationships between individuals’ characteristics and the treatments’ effects. This study proposes an ML method for estimating HTE based on the rule ensemble method termed RuleFit. The main advantages of RuleFit are interpretability and accuracy. However, HTEs are always defined in the potential outcome framework, and RuleFit cannot be applied directly. Thus, we modified RuleFit and proposed a method to estimate HTEs that directly interpret the interrelationships among the individuals’ features from the model. The actual data from the HIV study, ACTG 175 dataset, was used to illustrate the interpretation based on the ensemble of rules created by the proposed method. In addition, the numerical results confirm that the proposed method has high prediction accuracy compared to the previous method, indicating that the proposed method establishes an interpretable model with sufficient prediction accuracy.
\end{abstract}

\keywords{Rules ensemble\and heterogeneous treatment effect\and potential outcome\and interpretablility}

\section{Introduction}\label{sec1}

Randomized clinical trials (RCTs) are the gold standard for providing evidence for new drugs and treatments \citep{Hariton2018}. However, because they focus on specific issues, such as establishing efficacy and safety compared to standard treatments, and are conducted in particular populations, they do not fully capture the treatment effects that differ due to heterogeneity in the treatment population \citep{Bica2021}. With increasing interest in precision medicine, recent medical studies have indicated that the information about treatment in populations with heterogeneous characteristics in clinical trial data is incomplete, which has led to an increased emphasis on real-world data \citep{Agarwala2018, Rudrapatna2020}. Compared with RCT data, real-world data are observational data collected from routine medical practice through various sources, including electronic medical records, product registries, disease registries, etc. \citep{Klonoff2020}. Therefore, in such data, the treatment population is not specified and can provide complementary evidence to RCTs \citep{Wendling2018, Kim2018}. Real-world data contain information about the heterogeneity of patients and their outcomes after treatment. Therefore, it provides opportunities to detect the relationship between the effect of treatment and the heterogeneity of patients, to support clinical decision-making. However, it is also challenging to identify such relationships since real-world data are always complex, noisy, and high-dimensional. Therefore, recent studies into the heterogeneity of treatment effect (HTE) have proposed machine learning (ML) methods based on their superior performance on increasingly complex and high-dimensional data \citep{Gong2021}. However, the ML model is always a black box and this hampers interpreting the relationship between the characteristics of the individual and the treatment effect.

HTE refers to the causal effect of treatment or intervention on the outcome of interest for different individuals, defined in the potential outcome framework \citep{Neyman1990,Rubin1974}. For individual $i = 1,\cdots, N$, let $T_i\in\{0, 1\}$ be the binary treatment indicator for individual $i$, with $T_i = 0$ if individual $i$ is assigned to the control group and $T_i = 1$ if individual $i$ is assigned to the treatment group, and let $\bm{X}_i = (X_{i1}, \cdots, X_{ip})^T$ be the $p$-dimensional vector of covariates for individual $i$. The potential outcomes consist of $Y_i^{(0)}$ denoting the outcome for individual $i$ in the control group, and $Y_i^{(1)}$ denoting the outcome for individual $i$ in the treatment group. The observed outcomes can be denoted as: 
\begin{align*}
Y_i = T_iY_i^{(1)} + (1 - T_i) Y_i^{(0)}.
\end{align*}
Given $\{Y_i, T_i, \bm{X}_i\}_{i =1}^N$, the HTE is defined as 
\begin{align}
\label{hte}
\tau(\bm{x}_i) = \mathbb{E}(Y_i^{(1)} - Y_i^{(0)}|\bm{X}_i = \bm{x}_i),
\end{align}
where $\bm{x}_i$ is the observed $p$-dimensional vector of individual $i$. Therefore, the goal of HTE estimation is to estimate $\tau(\bm{x}_i)$. For individual $i$, the outcome is $Y_i = Y_i^{(1)}$ if $T_i = 1$, and the outcome is $Y_i = Y_i^{(0)}$ if $T_i = 0$. Therefore,  the HTE $\tau(\bm{x}_i)$ must be estimated because only one of the potential outcomes can be observed directly from the data and $\tau(\bm{x}_i)$ cannot be calculated directly. The general approach to dealing with such difficulties is to assume unconfoundedness \citep{Rosenbaum1983} as 
\begin{align}
\label{unconf}
\{Y_i^{(0)}, Y_i^{(1)}\} \indep T_i | \bm{X}_i,
\end{align} 
indicating that the treatment assignment $T_i$ is independent of the potential outcomes $Y_i^{(0)}$ and $Y_i^{(1)}$ conditions on $\bm{X}_i$. The conditional mean function for the control group $\mu^{(0)}(\bm{x})$ and treatment group $\mu^{(1)}(\bm{x})$ 
are denoted as 
\begin{align*}
\mu^{(0)}(\bm{x}_i) &= \mathbb{E}(Y_i| T_i = 0, \bm{X}_i = \bm{x}_i), \\
\mu^{(1)}(\bm{x}_i) &= \mathbb{E}(Y_i| T_i = 1, \bm{X}_i = \bm{x}_i)  
\end{align*}
Under the assumption of unconfoundedness (\ref{unconf}), the HTE (Eq.\ref{hte}) can be rewritten as 
\begin{align*}
\tau(\bm{x}_i) = \mu^{(1)}(\bm{x}_i) - \mu^{(0)}(\bm{x}_i).
\end{align*}

The approaches for estimating the HTE generally fall into two frameworks: transformed outcome regression and conditional mean regression~\citep{Powers2018,Wendling2018, Zhu2020}. Transformed outcome regression approaches build the HTE estimation model based on transformed outcomes. The transformed outcome $\tilde{Y}_i$ is the outcome weighted by the inverse “probability of treatment” weighting (IPTW) \citep{Lunceford2004} and can be described as follows: 
\begin{align*}
\tilde{Y}_i  = \frac{T_i}{\pi(\bm{x}_i)} Y_i - \frac{(1 - T_i)}{1-\pi(\bm{x}_i)} Y_i
\end{align*}
where
\begin{align*}
\pi(\bm{x}_i) = \mathrm{Pr} (T_i = 1| \bm{X}_i = \bm{x}_i)
\end{align*}
is the propensity score, which is the “probability of receiving treatment” condition in $\bm{x}_i$. The transformed outcome $\tilde{Y}_i$ has the property that it is the unbiased estimator of the treatment effect for an individual $i$. Therefore, within the framework of the transformed outcome regression approach, existing regression approaches can be applied directly to build an HTE estimation model. However, the transformed outcome regression approach exhibits weakness due to its high variance, whereas the propensity score $\pi(\bm{x}_i)$ approximates either 0 or 1 \citep{Michael}. Conditional mean regression approaches model the difference between the conditional mean of the control and treatment groups. Because variances can be large in transformed outcome regression approaches, conditional mean regression methods have been preferred in recent studies \citep{Athey2016,Powers2018}.

The transform outcome regression approach can be easily performed using previous methods. Nonetheless, its prediction accuracy is always inferior to that of the conditional mean regression approaches because of the potentially high variance of the transformed outcome estimator. \cite{Powers2018} pointed out the weakness of transformed outcome regression and proposed the “pollination” procedure in their study. The “pollination” procedure builds the base function in the framework of transformed outcome regression and then estimates the HTE based on these base functions in the framework of conditional mean regression. This procedure can maintain the easy usage of existing supervised methods and effectively improve the prediction accuracy of the transformed outcome regression.

In this study, we focused on estimating HTE using an ML method with an interpretable model to capture the relationship between the characteristics of the individual and the effect of treatment. However, the models of most previous ML methods are black boxes, making the relationship between the characteristics of the individual and the treatment effect challenging to interpret. To overcome this weakness, we focus on the rule ensemble method, RuleFit \citep{Friedman2008}. This method provides an interpretable rule-based ensemble model and has shown a prediction accuracy similar to the random forest and gradient boosting tree algorithms. In addition, we focus on the “pollination” procedure \citep{Powers2018} for easy use of existing supervised methods. Therefore, we propose an ML method for HTE estimation by RuleFit using the “pollination” procedure. To demonstrate the usefulness of the proposed method, we provide a numerical simulation to demonstrate its prediction accuracy. Furthermore, we apply it to real clinical data to demonstrate the interpretability of the proposed method.

\section{Related work}\label{sec2}

In recent studies, the ML methods for HTE estimation have been gaining attention for their excellent performance on complex and high-dimensional data. These methods generally fall into two frameworks: conditional mean regression and transformed outcome regression.

In the transformed outcome regression framework, the HTE prediction model is built based on the transformed outcome. The transformed outcome is the outcome weighted by IPTW \citep{Lunceford2004} and provides unbiased HTE estimators. Thus, any existing supervised learning method can be directly applied to transformed outcomes to build a model for HTE. For example, \cite{Athey2016} mentioned the transformed outcome tree (TOT), which applies a decision tree (CART) \citep{Breiman2017} directly to the transformed outcome.

In conditional mean regression frameworks, the HTE prediction model is built based on the difference between the conditional mean of the control and treatment groups. A causal tree \citep{Athey2016} performs a recursive partition that divides the data into subgroups with different treatment effects. It has also been developed into a causal forest \citep{Wager2018} based on random forest \citep{Breiman2001}. \cite{Athey2019} proposed the generalized random forest and provide another causal forest implementation procedure. \cite{Powers2018} proposed PTO forest, causal boosting, and bagged causal MARS for estimating HTE. Sugasawa et al. (2019) \cite{Sugasawa2019} proposed a method based on the gradient boosting tree (GBT) \citep{Friedman2001}. Bayesian approaches are also used to identify naturally the heterogeneous treatment effect \citep{Hill2011,Henderson2016}. \cite{Hill2011} proposed a strategy for causal inference using Bayesian additive regression trees (BART) \citep{Chipman2010}. Furthermore, \cite{Hahn2020} extend the ideas of \cite{Hill2011} and \cite{Hahn2018} and proposed the Bayesian causal forest (BCF) for estimating HTE when the effect size was small, treatment effect was heterogeneous, and strongly confounded by the observers.

These methods have demonstrated exemplary performance in estimating HTE. In particular, ensemble methods, such as causal forest and BART, have been the focus of recent studies because of their high prediction accuracy for complex and high-dimensional data. However, in contrast to its high prediction accuracy, the model constructed by ensemble learning is often a “black box,” which makes the results difficult to interpret. For such problems, a common approach is to build a new interpretable model based on the estimated HTE. \cite{Logan2019} proposed the "fit-to-fit" method which uses a single decision tree to interpret the HTE estimated by BART. \cite{Lee2020} proposed the Causal Rule Ensemble (CRE) method, which applies a rule ensemble method to HTE estimated by existing methods (BCF is recommended) to obtain an interpretable HTE model.

\section{Rule ensemble for HTE}\label{sec3}

As mentioned above, ensemble learning is often used to estimate HTEs because of its high prediction accuracy. However, this interpretation is difficult because these methods often render the model a “black box. Therefore, in this study, our objective is to propose an interpretable estimation approach using ensemble learning. This work was inspired by RuleFit \citep{Friedman2008}, which is a rule-based ensemble learning approach that uses the conjunction of several indicator functions as the base function. Due to the simple structure of each base function, the model can be easily interpreted. Therefore, we extend the RuleFit to the HTE estimation method. 

\subsection{RuleFit}

RuleFit, a rule-based ensemble method proposed by \cite{Friedman2008} has the main advantage of easy interpretability because its model is built using a linear combination of rule terms and linear terms. Given the covariate vectors $\bm{x} = (x_1, x_2, \cdots, x_p)^T \in \mathbb{R}^p$, the RuleFit model is defined as: 
\begin{align*}
F_{RFit}(\bm{x}) = \alpha_0 + \sum_{k=1}^K\alpha_kr_k(\bm{x}) + \sum_{j=1}^p\alpha^*_jl_j(x_j)
\end{align*} 
where $\alpha_0 \in \mathbb{R}$, $\alpha_k \in \mathbb{R} (k = 0, 1, \cdots, K)$ and $\alpha^*_j \in \mathbb{R} (j = 1, \cdots, p)$ are the intercept, coefficients of the rule terms, and coefficients of the linear terms, respectively. The rule terms consist of the conjunction of indicator functions $r_k : \mathbb{R}^p \mapsto \mathbb{R}$ and the linear terms consist of the function $l_j : \mathbb{R} \mapsto \mathbb{R}$. For the function of rule terms $r_k$ and linear terms $l_j$ are explained in detail as follows.

\paragraph{Rule terms}
Let $S_j$ be the set of all possible values of $j$th ($j = 1, \cdots, p$) variable $x_j$. Given $x_j \in S_j$, the function of $k$th rule term can be defined as
\begin{align}
\label{rule}
r_k(\bm{x}) = \prod_{j=1}^p I (x_j \in S_{jk})
\end{align} 
where $S_{jk} \subset S_j$ is the specified subset, $I(\cdot)$ is the indicator function that returns $1$ if the rule within the parentheses is true, and $0$ if it is false. Here, subset $S_{jk}$ is denoted as the interval.
\begin{align*}
S_{jk} = [x^-_{jk}, x^+_{jk})
\end{align*}
where $x^-_{jk}$ and $x^+_{jk}$ are the lower and upper bounds of $x_j$ in rule $r_{k}(\bm{x})$. Although the rule terms facilitate interpretation, the model based only on the rule terms performs poorly for the linear structure. To overcome this difficulty, RuleFit consists of both the rule and linear terms. 

\paragraph{Linear terms}
Although the model including the linear terms can improve the performance of a linear structure, it may reduce the robustness against outliers compared to the model based only on the rule terms. Therefore, the "winsorized" version of the linear term 
\begin{align}
l_j(x_j) = \min (\delta_j^+, \max(\delta_j^-, x_j)) \label{wins}
\end{align}
are used, where $\delta _j^+$ and $\delta_j^-$ are the $q$ and $(1-q)$ quantiles of $x_j$, respectively, and $q$ = 0.025 is recommended \citep{Friedman2008}. $\delta _j^+$ and $\delta_j^-$ are the thresholds for determining outliers. To ensure that the linear terms have the same prior influence as the rule terms, the linear term is normalized as: 
\begin{align}
l_j(x_j) \leftarrow 0.4\cdot l_j(x_j) / \mathrm{std} (l_j(x_j))\label{norms}
\end{align}
where $0.4$ is the average standard deviation of the rule term $r_k(\bm{x})$ under the assumption that the support of the rule terms from the training data
\begin{align}
\label{support}
\varrho_k = \frac{1}{N}\sum_{i=1}^N r_k(\bm{x}_i) 
\end{align}
is distributed uniformly $U(0,1)$. $\textrm{std} (l_j(x_j))$ is the standard deviation of $l_i(\bm{x}_j)$ for training data.

\subsection{Proposed method}

RuleFit builds the model as a linear combination of rule terms and linear terms; therefore, ease of interpretability is its main advantage. In addition, RuleFit has shown a prediction accuracy similar to that of random forest and gradient boosting tree \citep{Friedman2008}. Therefore, applying RuleFit for HTE estimation can be expected to obtain an interpretable model for HTE and ensure prediction accuracy comparable to that of previous tree-based ensemble methods, such as causal forest, BART, and PTO forest. 

\begin{figure}[t]
 \centering
 \includegraphics[width = \linewidth]{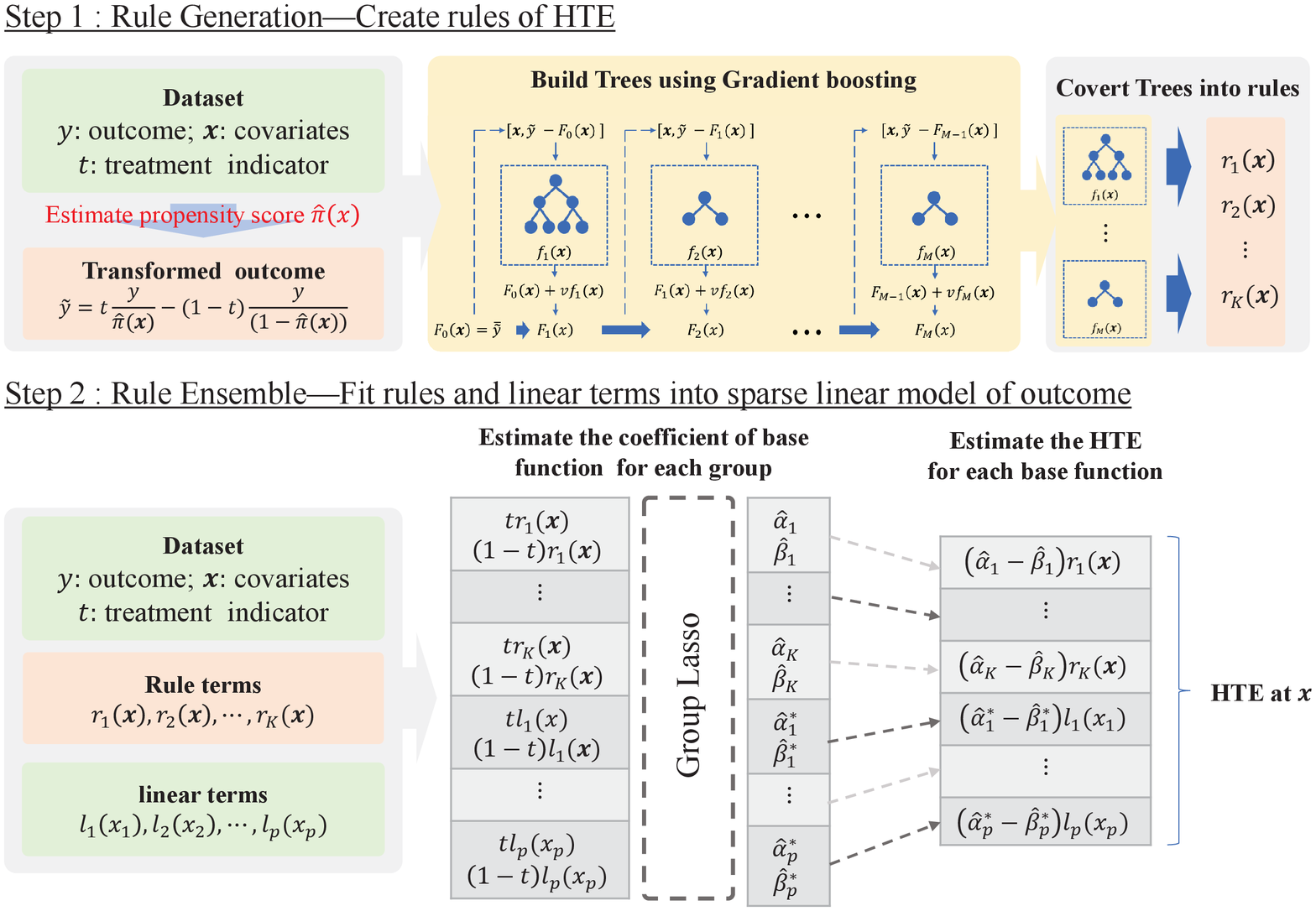}
 \caption{Summarized algorithm of the proposed method.}
 \label{HTE_algo}
\end{figure}

In this study, we have applied RuleFit for HTE estimation based on conditional mean regression, because the high variance of the transformed outcomes always makes it difficult to fit the model and leads to poor prediction accuracy. Given the dataset $\{y_i,t_i,\bm{x}_i\}_{i=1}^N$, where $y_i$ is the observed outcome for individual $i$, $t_i\in \{0, 1\}$ is the observed treatment indicator with $t_i = 0$ if individual $i$ is assigned to the control group, $t_i = 1$ if individual $i$ is assigned to the treatment group, and $\bm{x}_i = (x_{i1},x_{i2},\cdots,x_{ip})^T$ are the observed variables for individual $i$. The RuleFit models $F_{RFit}^{(1)}(\bm{x})$ for the treatment group ($t=1$) and $F_{RFit}^{(0)}(\bm{x})$ for the treatment group ($t=0$) must be constructed as follows:
\begin{align*}
F_{RFit}^{(0)}(\bm{x}) &= \alpha_0 + \sum_{k=1}^K\alpha_kr_k(\bm{x}) + \sum_{j=1}^p\alpha^*_jl_j(x_j) \quad \mathrm{and} \\
F_{RFit}^{(1)}(\bm{x}) &= \beta_0 + \sum_{k=1}^K\beta_kr_k(\bm{x}) + \sum_{j=1}^p\beta^*_jl_j(x_j)
\end{align*} 
where $\alpha_0, \beta_0 \in \mathbb{R}$, $\alpha_k, \beta_k \in \mathbb{R} (k = 0, 1, \cdots, K)$ and $\alpha^*_j,\beta^*_j \in \mathbb{R} (j = 1, \cdots, p)$ are the intercept, coefficients of the rule terms, and the coefficients of the linear terms, respectively. We combine these two models $F_{RFit}^{(0)}(\bm{x})$ and $F_{RFit}^{(1)}(\bm{x})$ and propose the model $F(\bm{x},t)$ of $y$, denoted as: 
\begin{align}
\label{propmodel}
F(\bm{x},t) = \theta_0 + t \Biggr[\sum_{k = 1}^K \alpha_k r_k (\bm{x}) + \sum_{j = 1}^p \alpha^*_j l_j(x_j)\Biggr] + (1-t)\Biggr[\sum_{k = 1}^K \beta_k r_k (\bm{x}) + \sum_{j = 1}^p \beta^*_j l_j(x_j)\Biggr]
\end{align}
where $\theta_0$ denotes the intercept. The model for the control and treatment groups should contain the same base function $r_k(x)$ and $l_j(x_j)$ because a different base function could lead to an incorrect HTE estimation \citep{Powers2018} and cannot obtain the base functions based on the HTE, reducing the interpretability of the HTE model. Therefore, our proposed method constrains the treatment and control group models to the same base functions, and the HTE can be calculated by the difference between $F(\bm{x},t = 1)$ and $F(\bm{x},t = 0)$ as:
\begin{align}
\label{HTE_est}
\mathrm{HTE} :\tau(\bm{x}) &= F(\bm{x},t = 1) - F(\bm{x},t = 0) \notag \\ 
                   &= \sum_{k=1}^K(\alpha_k - \beta_k)r_k(\bm{x}) + \sum_{j=1}^p(\alpha^*_j-\beta^*_j)l_j(x_j)
\end{align}
where $(\alpha_k - \beta_k)$ and $(\alpha^*_j-\beta^*_j)$ can be interpreted as the HTE based on rule terms and linear terms, respectively. Therefore, the proposed method calculates Eqs.(\ref{HTE_est}) under the estimated rule terms $r_k(\bm{x})$ and parameters $\alpha_k$, $\beta_k$, $\alpha^*_j$, and $\beta^*_j$ in Eq.(\ref{propmodel}). We propose an algorithm that modifies the original RuleFit algorithm based on the pollination procedure \citep{Powers2018}. This consists of two steps: rule generation and rule ensembling.    
\begin{description}
\item{\textbf{Step 1} Rule generation:} In this step, the rules of HTE were created based on the transformed outcome approach. Therefore, the transformed outcomes were calculated from the data, and a gradient boosting tree (GBT) \citep{Friedman2001} was fitted to the transformed outcomes. Then, we converted all tree-based learners of the fitted GBT model to rules and obtained many rule terms for HTE. 
\item{\textbf{Step 2} Rule ensembling:} In this step, the intercept and coefficients of rule terms and linear terms are estimated. We grouped the rule terms and the linear terms for the treatment and control group and used the group lasso to estimate these parameters. 
\end{description}
Our proposed method differs from the original RuleFit in two ways. First, rules were created based on the transformed outcome rather than the outcome. Second, the original RuleFit fits the rule terms and linear terms into a sparse linear model using a least absolute shrinkage and selection operator (lasso) \citep{Tibshirani1996}. However, we constrained the same base functions for the treatment group ($t=1$) and control group ($t=0$) models. Differently expressed, a group of parameters must be estimated for each base function. Therefore, instead of lasso, we use the group lasso, which can spark our base function to be sparsed at the group level. We summarize the algorithm of the proposed method in Fig.\ref{HTE_algo}. The algorithm is illustrated in detail in the next subsection.

\subsubsection{Rule generation}

In this step, the rules of the proposed model are constructed. First, the GBT model is used to transform the outcome and then convert all the regression trees in the constructed GBT model into rules to obtain the HTE rule terms . Therefore, given the dataset $\{y_i,t_i,\bm{x}_i\}_{i = 1}^n, \bm{x}_i = (x_{i1},x_{i2},\cdots, x_{ip})^T$, the estimated propensity score $\hat{\pi}_i(\bm{x}_i) = \Pr(t_i=1|\bm{x}_i)$ (i.e. It can simply be $\hat{\pi}_i(\bm{x}_i) = 0.5$ in RCT data), and the transformed outcome for the $i$ th individual is calculated as
\begin{align}
    \tilde{y}_i \leftarrow t_i\frac{y_i}{\hat{\pi}(\bm{x}_i)} + (1-t_i)\frac{-y_i}{1-\hat{\pi}(\bm{x}_i)} .
\end{align}
Then the GBT model $F(\bm{x}_i)$ of $\tilde{y}_i$ is built in the form of $F(\bm{x}_i) = \sum_{m=1}^M f_m(\bm{x}_i)$ where $M$ is the number of regression trees within the GBT model and $f_m(\bm{x}_i) = \sum_{t=1}^{t_m} \gamma_{tm}I(\bm{x}_i\in R_{tm})$ represents the regression tree $m$-th ($m = 1,2,\cdots,M$) with disjoint partitioned regions $R_{1m}, R_{2m}, \cdots, R_{t_mm}$ and $\gamma_{t_m}$ is the prediction value for the corresponding region $R_{t_m}$. Then, the whole-tree model $\{f_m(\bm{x}_i)\}_{m=1}^M$ is converted into rules. The rule generation algorithm is detailed in algorithm\ref{Rulefit_create}.

\begin{algorithm}[tb]
\caption{Rule generation}
\label{Rulefit_create}
\begin{algorithmic}[1]
\STATE \textbf{Input} : Dataset $\{y_i, t_i, \bm{x}_i\}_{i = 1}^N$, number of tree-based learners $M$, mean depth of tree-based learners $\bar{L}$, shrinkage rate $v$, and the training sample fraction for each tree-based learner $\eta$
\STATE For $i = 1, 2, \cdots, N$, compute the transformed outcome $\tilde{y}_i$
\begin{align*}
\tilde{y}_i \leftarrow t_i\frac{y_i}{\hat{\pi}(x_i)} + (1-t_i)\frac{-y_i}{1-\hat{\pi}(x_i)} 
\end{align*}
\STATE Initialize the model $F_0(\bm{x}_i) \leftarrow \bar{\tilde{y}}$
\FOR{$m = 1$ to $M$}
\STATE For $i = 1, 2, \cdots, N$, compute the pseudo residual
\begin{align*}
z_{im} \leftarrow \tilde{y}_i - F_{m-1}(\bm{x}_i)
\end{align*}
\STATE Determine the number of terminal nodes for the tree-based learner, $t_m = 2 + \mathrm{floor}(u)$, where $u \sim \mathrm{exp}(1/(\bar{L} - 2))$
\STATE Fit a regression tree to the pseudo residual $z_{im}$, giving the terminal regions $R_{t_m}, t = 1, 2, \cdots t_m$
\STATE For $t = 1,2,\cdots,t_m$, estimate the value of region $R_{t_m}$
\begin{align*}
\hat{\gamma}_{t_m} = \argmin_{\gamma_{t_m}}\sum_{\bm{x}_i \in R_{t_m}}\sum_{i \in H} (z_{im} - F_{m-1}(\bm{x}_i) - \gamma_{t_m})^2
\end{align*}
where $H \subset \{1,2,\cdots, N\}$ and $|H| = \lfloor \eta N \rfloor$ \\
\STATE Update $F_{m}{(\bm{x}_i)} \leftarrow F_{m-1}(\bm{x}_i) + v\cdot \sum_{t=1}^{t_m}\hat{\gamma}_{t_m} I(\bm{x}_i \in R_{t_m}) = F_{m-1}(\bm{x}) + vf_m(\bm{x})$
\ENDFOR
\STATE For $m = 1,2,\cdots, M$, traverse the tree $f_m(\bm{x})$ to compile the rule set $\{r_{km}(\bm{x})\}_{k=1}^{K_m}, K_m = 2(t_m - 1)$ 
\STATE Collect every rule set $\{r_{k1}(\bm{x})\}_{k=1}^{K_1}, \{r_{k2}(\bm{x})\}_{k=1}^{K_2},\cdots ,$ and $\{r_{kM}(\bm{x})\}_{k=1}^{K_M}$ into $\{r_{k}(\bm{x})\}_{k=1}^{K}$, where $K = \sum_{m =1}^M K_m$.
\STATE \textbf{Output} $\{r_{k}(\bm{x})\}_{k=1}^{K}$
\end{algorithmic}
\end{algorithm}

The data $\{y_i, t_i, \bm{x}_i\}_{i = 1}^N$, number of tree-based learners $M$, mean depth of the tree-based learners $\bar{L}$, the shrinkage rate $v$, and the training sample fraction for each tree-based learner $\eta$ are the inputs. The rule generation step first calculates the transformed outcomes from the data, as shown in line 2. Then, the GBT model $F_M(\bm{x}_i)$ with $M$ trees is built for the transformed outcome $\tilde{y}_i$ as in lines 3 to 10. In line 3, the GBT model is initialized. For $m = 1, 2, \cdots, M$, the pseudo-outcome $z_m$ is calculated as line 5. Then, as in lines 6 to 8, tree-based learner $f_m(\bm{x}_i)$ of the pseudo-outcome $z_m$ is built, and it can be simply performed using the R package $\texttt{rpart}$ \citep{rpart}. Here, as in line 6, it should be noted that the depth of the trees is randomly determined, where $\mathrm{floor}(u)$ is the largest integer less than or equal to $u$ and $\mathrm{exp}()$ denotes the exponential distribution. This is a key process for RuleFit that allows us to obtain rules with various depths of interaction. Next, as in line 9, we update the model $F_m(\bm{x}_i)$, where $v$ is the shrinkage rate that takes a small value. Finally, all trees from the constructed GBT models are converted into rules, as shown in lines 11 and 12. For the GBT model containing $M$ trees, all trees $\{f_m(\bm{x}_i)\}_{m= 1}^M$ are decomposed into $K$ rules $\{r_{k}(\bm{x}_i)\}_{k = 1}^K$, where the number of rules is
\begin{align*}
K = \sum_{m = 1}^M 2(t_m - 1)
\end{align*}
where $t_m$ denotes the number of terminal nodes in the $m$th tree. 

\subsubsection{Rule ensemble}

The main objective of this step is to estimate the $\theta_0$, $\alpha_k, \beta_k, \alpha^*_j$, and $\beta^*_j$ parameters in Eq.(\ref{propmodel}) by fitting the rule terms and the "winsorized" version of the linear terms to the sparse linear models using the group lasso. Because the rule terms $\{r_k(\bm{x})\}_{k=1}^K$ are generated in the rule generation step, the "winsorized" version of the linear terms is calculated. Given the data $\{y_i, t_i, \bm{x}_i\}_{i = 1}^N$ the "winsorized" version of linear terms is denoted as 
\begin{align*}
l_j(x_j) = \min (\delta_j^+, \max(\delta_j^-, x_j))
\end{align*}
where $\delta_j^+$ and $\delta_j^-$ are the $q$ and $(1-q)$ quantiles of $\{x_{ij}\}_{j = 1}^N, j = 1,2,\cdots,p$. Similar to the original RuleFit, to ensure that the linear terms have the same selection priority as the rule terms in the group lasso, the linear terms are normalized as: 
\begin{align*}
    l_j(x_j) \leftarrow 0.4\frac{l_j(x_j)}{std(l_j(x_j))}, j = 1,2,\cdots,p.
\end{align*}

Then, the rule terms $\{r_k(\bm{x})\}_{k=1}^K$ and the "winsorized" version of the linear terms $\{l_j(x_j)\}_{j=1}^p$ are fitted to the sparse linear models using the group lasso. We first introduce the group lasso and then demonstrate its use in our proposed method.

\paragraph{Group lasso} This method is the extension of lasso, which allows variable selection on a predefined group of variables in linear regression models \citep{Yuan2006}. Given data $\{y_i,\bm{g}_{i1},\bm{g}_{i2},\cdots, \bm{g}_{iJ}\}_{i=1}^N$ consisting of $J$ groups of variables, where the grouped variable $\bm{g}_j\in \mathbb{R}^{p_j} (j = 1,2,\cdots,J)$ and $p_j$ is the number of variables in group $j$. The estimated intercept $\theta_0 \in \mathbb{R}$ and coefficients of the grouped variable $\bm{\theta}_j\in\mathbb{R}^{p_j} (j=1,2,\cdots,J)$ are then defined as: 
\begin{align*}
(\hat{\theta}_0,\{\hat{\bm{\theta}}_j\}_{j=1}^J) = \argmin_{\theta_0,\{\bm{\theta}_j\}_{j=1}^J} \frac{1}{2}\sum_{i=1}^N (y_i - \theta_0 - \sum_{j = 1}^J {\bm{g}_{ij}}^T\bm{\theta}_j)^2 + \lambda\sum_{j = 1}^J\sqrt{p_j}||\bm{\theta}_j||_2
\end{align*}
where $||\bm{\theta}_j||_2$ is the Euclidean norm of the coefficient vector $\bm{\theta}_j$ and $\lambda \geq 0$ is a tuning parameter that can be optimized by cross-validation. The group lasso can be implemented rapidly in the R package $\texttt{grpreg}$ \citep{grpreg}.

\paragraph{Application of the group lasso in the proposed method} To estimate the parameters  $\alpha_k, \alpha^*_j$ of the treatment group model and $\beta_k, \beta^*_k$ of the control group model simultaneously using the group lasso, we define the grouped variables vectors $\bm{z}_k$ and $\bm{z}_j$ based on the estimated rule terms. We calculate the “winsorized” version of linear terms in Eq.(\ref{propmodel}) as follows: 
\begin{align}
\bm{z}_{ik} &= \left(t_ir_k(\bm{x}_i), (1-t_i)r_k(\bm{x}_i)\right)^{T}, k = 1,2,\cdots,K  \quad \mathrm{and} \label{group_rule}\\
\bm{z}^*_{ij} &= \left(t_il_j(x_{ij}), (1-t_i)l_j(x_{ij})\right)^{T}, j = 1,2,\cdots,p; i = 1,2,\cdots,N \label{group_linear}
\end{align}
respectively. 
Then, the original data can be transformed into the form of $\{y_i,\bm{z}_{i1},\bm{z}_{i2}\cdots\,\bm{z}_{iK},\bm{z}^*_{i1},\bm{z}^*_{i2}\cdots\,\bm{z}^*_{ip}\}_{i=1}^N$ and the group lasso can be applied to estimate the parameters. 

We then use the group lasso to estimate the parameters of the objective function based on the value of the grouped rule terms $\bm{z}_{ik} (k = 1,2,\cdots,K)$ (Eq.\ref{group_rule}) and the grouped linear terms $\bm{z}_{ij} (j = 1,2,\cdots,p)$ (Eq.\ref{group_linear}), thus the group lasso is applied to the data $\{y_i,\bm{z}_{i1},\bm{z}_{i2}\cdots\,\bm{z}_{iK},\bm{z}^*_{i1},\bm{z}^*_{i2}\cdots\,\bm{z}^*_{ip}\}_{i=1}^N$, and the parameter can be estimated as
\begin{align}
(\hat{\theta}_0,\{\hat{\bm{\theta}}_k\}_{k=1}^K, \{\hat{\bm{\theta}}^*_j\}_{j=1}^p) &= \argmin_{\theta_0,\{\bm{\theta}_k\}_{k=1}^K,\{\bm{\theta}^*_j\}_{j=1}^p} \frac{1}{2}\sum_{i=1}^N (y_i - \theta_0 - \sum_{k = 1}^K {\bm{z}_{ik}}^T\bm{\theta}_k - \sum_{j = 1}^p {\bm{z}^*_{ij}}^T\bm{\theta}^*_j)^2 \notag \\
& + \lambda\sqrt{2}\left(\sum_{k = 1}^K||\bm{\theta}_k||_2 + \sum_{j = 1}^p||\bm{\theta}^{*}_j||_2\right)
\end{align}
where $\hat{\bm{\theta}}_k = (\hat{\alpha}_k, \hat{\beta}_k)^T (k = 1,2,\cdots, K)$ is the estimated coefficient vector for $\bm{z}_{ik}$ and $\hat{\bm{\theta}}^*_j = (\hat{\alpha}^*_j, \hat{\beta}^*_j)^T (j = 1,2,\cdots, p)$ is the coefficient vector for $\bm{z}_{ij}$.

The algorithm \ref{Rulefit_ensemble} of the rule ensemble is presented in detail.
\begin{algorithm}[tb]
\caption{Rule ensemble}
\label{Rulefit_ensemble}
\begin{algorithmic}[1]
\STATE \textbf{Input}: Data $\{y_i, t_i, \bm{x}_i\}_{i = 1}^N$, estimated rules $\{r_k(\bm{x})\}_{k=1}^{K}$ and "winsorize" margin $\delta^-_j$ and $\delta^+_j$
\STATE For $j = 1, 2, \cdots,p$, obtain "winsorized" linear terms 
\begin{align*}
&l_j(x_j) = \min (\delta_j^+, \max(\delta_j^-, x_j))
\end{align*}
\STATE For $j = 1,2,\cdots,p$, normalize the $l_j(x_j)$ as
\begin{align*}
 l_j(x_j) \leftarrow 0.4\frac{l_j(x_j)}{std(l_j(x_j))}
\end{align*}
\STATE Define the grouped variables vectors based on the rule terms and linear terms in our proposed model Eq.(\ref{propmodel}).
\begin{align*}
\bm{z}_{ik} &= \left(t_ir_k(\bm{x}_i), (1-t_i)r_k(\bm{x}_i)\right)^{T}, k = 1,2,\cdots,K  \quad \mathrm{and} \notag \\
\bm{z}^*_{ij} &= \left(t_il_j(x_{ij}), (1-t_i)l_j(x_{ij})\right)^{T}, j = 1,2,\cdots,p; i = 1,2,\cdots,N \notag
\end{align*}
\STATE Obtaining the new data $\{y_i,\bm{z}_{i1},\bm{z}_{i2},\cdots,\bm{z}_{iK},\bm{z}^*_{i1},\bm{z}^*_{i2},\cdots,\bm{z}^*_{ip}\}_{i=1}^N$ consist of the predefined grouped variable
\STATE Estimating the coefficient vector of the grouped variable $\hat{\bm{\theta}}_k = (\hat{\alpha}_k, \hat{\beta}_k)^T (k = 1,2,\cdots, K)$ and $\hat{\bm{\theta}}^*_j = (\hat{\alpha}^*_j, \hat{\beta}^*_j)^T (j = 1,2,\cdots, p)$
\begin{align*}
(\hat{\theta}_0,\{\hat{\bm{\theta}}_k\}_{k=1}^K, \{\hat{\bm{\theta}}^*_j\}_{j=1}^p) &= \argmin_{\theta_0,\{\bm{\theta}_k\}_{k=1}^K,\{\bm{\theta}^*_j\}_{j=1}^p} \frac{1}{2}\sum_{i=1}^N (y_i - \theta_0 - \sum_{k = 1}^K {\bm{z}_{ik}}^T\bm{\theta}_k - \sum_{j = 1}^p {\bm{z}^*_{ij}}^T\bm{\theta}^*_j)^2 \notag \\
& + \lambda\left(\sum_{k = 1}^K\sqrt{2}||\bm{\theta}_k||_2 + \sum_{j = 1}^p\sqrt{2}||\bm{\theta}^*_j||_2\right)
\end{align*}
\STATE \textbf{Output} $\hat{\theta}_0,\{\hat{\alpha}_k\}_{k=1}^K, \{\hat{\beta}_k\}_{k=1}^K, \{\hat{\alpha}^*_j\}_{j=1}^p$ and $\{\hat{\beta}^*_j\}_{j=1}^p$
\end{algorithmic}
\end{algorithm}
Input the data $\{y_i, t_i, \bm{x}_i\}_{i = 1}^N$ the estimated rules $\{r_k(\bm{x})\}_{k=1}^K$ and "winsorized margin" $\delta^-_j$ and $\delta^+_j$. In the rule ensemble step, for $j = 1,2,\cdots,p$, as in line 2, we obtain the "winsorized" linear terms and normalize the linear terms as in line 3. Then, to estimate the parameters in Eq.(\ref{propmodel}) using the group lasso, we define the grouped variables as in line 4. In line 5, the data consisting of predefined grouped variables are created from the original data. Then, in line 6, the group lasso is used to estimate the intercept and coefficient of the rule and linear terms $\hat{\theta}_0,\{\hat{\alpha}_k\}_{k=1}^K, \{\hat{\beta}_k\}_{k=1}^K, \{\hat{\alpha}^*_j\}_{j=1}^p$ and $\{\hat{\beta}^*_j\}_{j=1}^p$ from the new data created in line 5.

\subsubsection{HTE estimation}

According to the rule generation and ensemble steps, the $\{r_k(\bm{x})\}_{k=1}^K$ rule terms and the $\hat{\theta}_0,\{\hat{\alpha}_k\}_{k=1}^K, \{\hat{\beta}_k\}_{k=1}^K, \{\hat{\alpha}^*_j\}_{j=1}^p$, and $\{\hat{\beta}^*_j\}_{j=1}^p$ parameters in Eq.(\ref{propmodel}) are estimated; therefore, the HTE can be estimated using Eq.(\ref{HTE_est}) as follows: 
\begin{align}
\label{HTE_est1}
\mathrm{HTE} :\tau(\bm{x}) &= F(\bm{x},t = 1) - F(\bm{x},t = 0) \notag \\ 
                   &= \sum_{k=1}^K(\hat{\alpha}_k - \hat{\beta}_k)r_k(\bm{x}) + \sum_{j=1}^p(\hat{\alpha}^*_j-\hat{\beta}^*_j)l_j(x_j)
\end{align}
Thus, the proposed method can estimate the HTE and detect the relationship between the HTE and personal characteristics based on the constructed rules. Although the group lasso has been used to prune most rules unrelated to the HTE, with an increasing number of tree-based learners in the rule generation steps, too many rules could still be included in the final model, leading to a complex result. To manage this situation and facilitate interpretation and comprehension, the original RuleFit \citep{Friedman2008} provides a method to calculate the importance of rules and linear terms. These are used to measure the relevance to the HTE of the rules and linear terms in Eq.(\ref{HTE_est}). The higher importance of rules and linear terms reflects that these are more related to the HTE and helps to focus attention on the more critical result rather than all the results. Here, we modified the importance of the rules and linear terms. Given the data $\{y_i,t_i,\bm{x}_i\}_{i=1}^N$, the set of estimated rule terms $\{r_k(\bm{x})\}_{k=1}^K$, the estimated model parameters $\{\hat{\alpha_k}\}_{k=1}^K, \{\hat{\beta_k}\}_{k=1}^K$ and $\{\hat{\alpha^*_j}\}_{j=1}^p$, the $k$-th rule terms importance $I_k$ and the $j$-th linear terms importance $I_j$ are denoted as: 
\begin{align}
    I_k &= |\hat{\alpha}_k - \hat{\beta}_k|\cdot\sqrt{\varrho_k(1-\varrho_k)} \quad \mathrm{and} \label{imp.rule}\\
    I_j &= |\hat{\alpha}^*_j - \hat{\beta}^*_j|\cdot|l_j(x_j)-\bar{l}_j| \label{imp.linear}
\end{align}
where $\varrho_k$ is the support for the rules, as shown in Eq.(\ref{support}).

\section{Simulation study}\label{sec4}

Several numerical simulations using artificial data were performed to evaluate the prediction accuracy of the proposed method under various conditions. The proposed method was compared with previous methods. First, we show the design of the artificial data for the simulation studies, then the application result for each simulation design.

\subsection{Simulation design}
Each simulation study used a dataset with $N = 600$ samples $\{y_i, t_i, \bm{x}_i\}_{i = 1}^N$. For each individual $i$, $y_i$ is a continuous value response variable, $\bm{x}_i = (x_{i1}, x_{i2}, \cdots ,x_{ip})^T$ are the explanatory variables, and the different numbers of variables are set as $p = 50, 100, 200, 400$. Furthermore, $t_i\in\{0, 1\}$ is the binary treatment indicator, where $t_i = 0$ indicates that subject $i$ is assigned to the control group, and $t_i = 1$ indicates that subject $i$ is assigned to the treatment group. The simulation design is described in detail as follows.  

\paragraph{Explanatory variables} Explanatory variables $\bm{x}_i = (x_{i1}, x_{i2}, \cdots ,x_{ip})^T$ include both continuous and binary variables, where the odd-numbered variables $x_{ij}^{odd} (j = 1, 3, 5, \cdots)$ are continuous values randomly drawn from the standard normal distribution $x_{ij}^{odd} \sim N(0, 1)$. The even-numbered variables $x_{ij}^{even} (j = 2, 4, 6, \cdots)$ are binary values randomly drawn from the Bernoulli distribution $x_{ij}^{even} \sim B(0.5)$.

\paragraph{Treatment Indicators} The treatment indicator $t_i\in\{0, 1\}$ is simulated for both the condition of the randomized clinical trial and the observational study. It is created from the Bernoulli distribution as $t_i \sim B(\pi(\bm{x}_i))$, where the function $\pi(\bm{x}_i)$ shows the probability of $t_i = 1$. Therefore, $\pi(\bm{x}_i)$ should be considered separately to obtain the condition of the randomized clinical trial and the observational study.  

First, in the randomized clinical trial we assumed that individual $i$ is randomly assigned to the control group or the treatment group; thus,  $\pi(\bm{x}_i) = 0.5$. 

Second, we assumed that treatment indicators were obtained from an observational study. Unlike a randomized clinical trial, the control and treatment groups are not strictly assigned at random; the assignment of the groups is always biased for various reasons. In this simulation, considering the condition that the patients with the higher covariates effect are more likely to receive the treatment and $\pi(\bm{x}_i) = [\exp(\mu(\bm{x}_i) - \tau(\bm{x}_i)/2)]/[1 + \exp(\mu(\bm{x}_i) - \tau(\bm{x}_i)/2)]$, where the functions $\mu(\bm{x}_i)$ and $\tau(\bm{x}_i)$ are the impact of covariates on outcome and the HTE at $\bm{x}_i$, respectively\citep{Powers2018}.

\paragraph{Outcome model} The continuous outcomes from following models are generated as: 
\begin{align}
\label{out_model}
y_i = \mu(\bm{x}_i) + \left(t_i - \frac{1}{2}\right)\tau(\bm{x}_i) + \epsilon_i
\end{align}
where the functions $\mu(\bm{x}_i)$ and $\tau(\bm{x}_i)$ respectively are the impacts of the covariate on the outcome and the HTE in $\bm{x}_i$ while $\epsilon_i \sim N(0, 0.25)$ are the noise terms. To evaluate the prediction accuracy of the proposed method on datasets with different covariates and outcome relationships, the outcome model was generated in 12 scenarios, as shown in Table \ref{HTE.model}. The candidate function $\mu(\bm{x}_i)$ includes linear, piecewise constant, and non-linear functions. The candidate function $\tau(\bm{x}_i)$ includes constant, linear, piecewise constant functions, as well as a function involving both quadratic and interaction. Then, 12 different outcome models were built using different combinations of the functions $\mu(\bm{x}_i)$ and $\tau(\bm{x}_i)$.
\begin{table}
\scriptsize
\caption{Summary of 12 different outcome models}\label{HTE.model}
\begin{tabular*}{\linewidth}{@{\extracolsep\fill}cll@{\extracolsep\fill}}
\toprule
\textbf{Scenario} & \textbf{Covariate effect}  & \textbf{HTE}    \\
\midrule
Scenario 1          &    $\mu_1(\bm{x}_i)=-0.25 + 0.5\sum_{j = 1}^3 x_{ij}$     & $\tau_1(\bm{x}_i) = 2$          \\
Scenario 2          &    $\mu_1(\bm{x}_i)=-0.25 + 0.5\sum_{j = 1}^3 x_{ij}$     & $\tau_2(\bm{x}_i) = x_{i1} + x_{i2} + x_{i3} -  x_{i4} + x_{i5}$     \\
Scenario 3          &    $\mu_1(\bm{x}_i)=-0.25 + 0.5\sum_{j = 1}^3 x_{ij}$     & $\tau_3(\bm{x}_i) = 2\sum_{j = 1}^5I(x_{ij} >0) -5$    \\
Scenario 4          &    $\mu_1(\bm{x}_i)=-0.25 + 0.5\sum_{j = 1}^3 x_{ij}$     & $\tau_4(\bm{x}_i) = \frac{1}{\sqrt{2}}(x_{i1}^2 + x_{i3}^2 + x_{i5}^2 + 4x_{i2}(1-x_{i4}) - 4) $     \\ \hline
Scenario 5          &    $\mu_2(\bm{x}_i)=0.7I(x_{i1} > -1) -1.4I(x_{i2} >0) + 0.7(x_{i3} >1)$           &  $\tau_1(\bm{x}_i) = 2$    \\
Scenario 6          &    $\mu_2(\bm{x}_i)=0.7I(x_{i1} > -1) -1.4I(x_{i2} >0) + 0.7(x_{i3} >1)$           &   $\tau_2(\bm{x}_i) = x_{i1} + x_{i2} + x_{i3} -  x_{i4} + x_{i5}$   \\
Scenario 7          &    $\mu_2(\bm{x}_i)=0.7I(x_{i1} > -1) -1.4I(x_{i2} >0) + 0.7(x_{i3} >1)$           &   $\tau_3(\bm{x}_i) = 2\sum_{j = 1}^5I(x_{ij} >0) -5$   \\
Scenario 8          &    $\mu_2(\bm{x}_i)=0.7I(x_{i1} > -1) -1.4I(x_{i2} >0) + 0.7(x_{i3} >1)$           &    $\tau_4(\bm{x}_i) = \frac{1}{\sqrt{2}}(x_{i1}^2 + x_{i3}^2 + x_{i5}^2 + 4x_{i2}(1-x_{i4}) - 4) $   \\ \hline
Scenario 9          &    $\mu_3(\bm{x}_i)=\sin^2(x_{i1} + x_{i3}) - 2x_{i2}\exp(-(x_{i4}-x_{i5})^2)$           &   $\tau_1(\bm{x}_i) = 2$   \\
Scenario 10         &   $\mu_3(\bm{x}_i)=\sin^2(x_{i1} + x_{i3}) - 2x_{i2}\exp(-(x_{i4}-x_{i5})^2)$          &  $\tau_2(\bm{x}_i) = x_{i1} + x_{i2} + x_{i3} -  x_{i4} + x_{i5}$    \\
Scenario 11         &   $\mu_3(\bm{x}_i)=\sin^2(x_{i1} + x_{i3}) - 2x_{i2}\exp(-(x_{i4}-x_{i5})^2)$            & $\tau_3(\bm{x}_i) = 2\sum_{j = 1}^5I(x_{ij} >0) -5$     \\
Scenario 12         &   $\mu_3(\bm{x}_i)=\sin^2(x_{i1} + x_{i3}) - 2x_{i2}\exp(-(x_{i4}-x_{i5})^2)$            & $\tau_4(\bm{x}_i) = \frac{1}{\sqrt{2}}(x_{i1}^2 + x_{i3}^2 + x_{i5}^2 + 4x_{i2}(1-x_{i4}) - 4) $      \\ \hline
\bottomrule
\end{tabular*}
\end{table}

\paragraph{Performance evaluation} To evaluate the performance of the proposed method, the training and test dataset were created in an identical setting and the proposed method’s performance was evaluated by its prediction accuracy using the mean squared error (MSE) as 
\begin{align}
MSE = \frac{1}{N}\sum_{i = 1}^N(\tau(\bm{x}_i) - \hat{\tau}(\bm{x}_i))^2
\end{align}
where $\tau(\bm{x}_i)$ is the true HTE and $\hat{\tau}(\bm{x}_i)$ is the estimated value. In addition, we also provide several previous ensemble methods, such as causal forest, BART, bagged causal Mars, and PTO forest, to compare their prediction accuracy with the proposed method. 

The simulation was performed using the \texttt{R 4.1.2.} programming environment and all previous methods can be performed using the existing \texttt{R} packages. Causal forest was implemented using the R package \texttt{grf} \citep{grf} with each hyperparameter tuned, BART was implemented using the R package \texttt{bartCause} \citep{bartCause} with its default hyperparameters, and the PTO forest and bagged causal MARS were implemented using the R package \texttt{causalLearning} \citep{causalLearning} with its default hyperparameters. The proposed method is based on RuleFit; therefore, we refer to the parameter setting in \cite{Friedman2008}, where the maximum number of tree-based learners is $M = 333$, the mean depth of each tree-based learners is $\bar{L} = 2$, and the shrinkage rate is $v = 0.01$. The fraction of the sample used to train the tree-based learner in each boost step is $\eta = \min(N/2, 100 + 6\sqrt{N})$, where $N$ is the size of the entire training sample. In addition, for all methods, the propensity score was estimated using BART \citep{Wendling2018}. 

\subsection{Simulation results}

\begin{figure}[t]
 \centering
 \includegraphics[width = \linewidth]{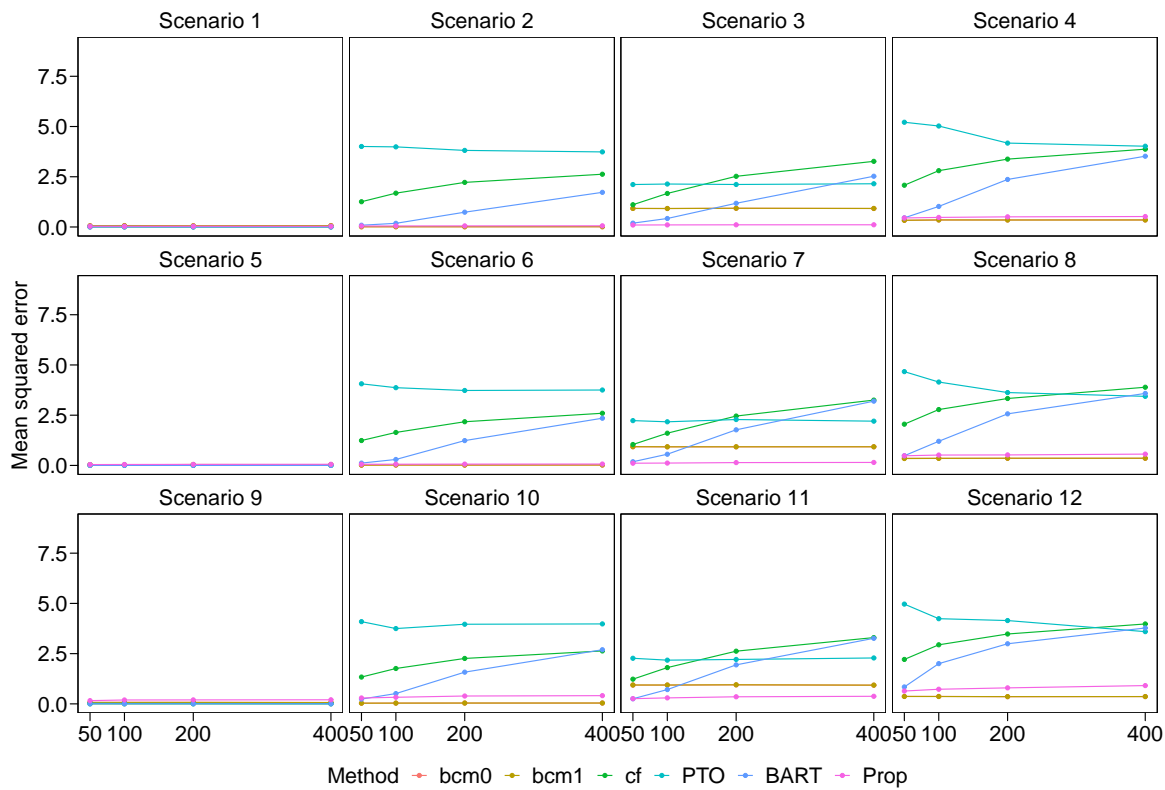}
 \caption{Results of simulated RCT datasets for twelve scenarios. 
For details of the generation model, see Table \ref{HTE.model}. The x-axis is the number of covariates in the simulation dataset for the corresponding scenario. The y-axis is the value of the mean squared errors (MSEs). These line charts show the median value of MSEs between the true and estimated HTEs with six different methods, including bcm0 = bagged causal MARS, bcm1 = bagged causal MARS, cf = causal forest, PTO = pollinated transformed outcome forest, BART = Bayesian additive regression tree, Prop = proposed method.}
 \label{rct_change}
\end{figure}

The results of the simulated RCT datasets for the twelve scenarios are shown in Fig.\ref{rct_change}. Each line chart shows the tendency of the median value of the MSE with an increasing number of covariates for the corresponding scenarios. The proposed method outperforms the other previous methods in Scenarios 3, 7, 11, and only BART shows prediction accuracy comparable to the proposed method when the number of covariates is small. Still, as the number of covariates increases, the prediction accuracy of BART tends to be poor. In these scenarios, the HTE is created from the step function, and the proposed method, which includes a rule term, can cover such a structure. Therefore, these situations are most suitable for the proposed method and obtained the best prediction accuracy. We then focus on scenarios 2, 6, and 10, where the HTE has linear relationships among the covariates. The bagged causal MARS performed best in these scenarios. The prediction accuracy of the proposed method is similar to that of the bagged causal MARS in scenarios 2 and 6 but a little less than that of the bagged causal MARS in scenario 10. Although the rule terms in the proposed model are not suitable for such linear structures, the linear terms in the model improve the prediction accuracy of the rule terms, and the performance is comparable to Bagged causal MARS, which has shown good performance for linear structures. In scenarios 4, 8, and 12, the HTE had quadratic and interaction relationships among the covariates. These scenarios are still suitable for bagged causal MARS, which performed the best under such conditions. Our proposed model cannot cover such structural HTEs, but its prediction accuracy tends to be poor. Finally, focusing on scenarios 1, 5, and 9, the HTE is constant; thus, these scenarios can be regarded as estimating the covariate impact on the outcome. All methods show similar results in scenarios 1 and 2; however, in scenario 9, the prediction accuracy of the proposed method is slightly inferior to that of previous methods. Thus, the complicated impact of non-linear covariates may perturb the prediction accuracy of the proposed method, which is also reflected in scenarios 10, 11, and 12. Furthermore, the results of the simulated RCT datasets of 12 scenarios show that the prediction accuracy of the proposed method and bagged causal MARS outperforms those of the causal forest, BART, and PTO forest. The prediction accuracy remains stable as the number of covariates increases.

\begin{figure}[t]
 \centering
 \includegraphics[width = \linewidth]{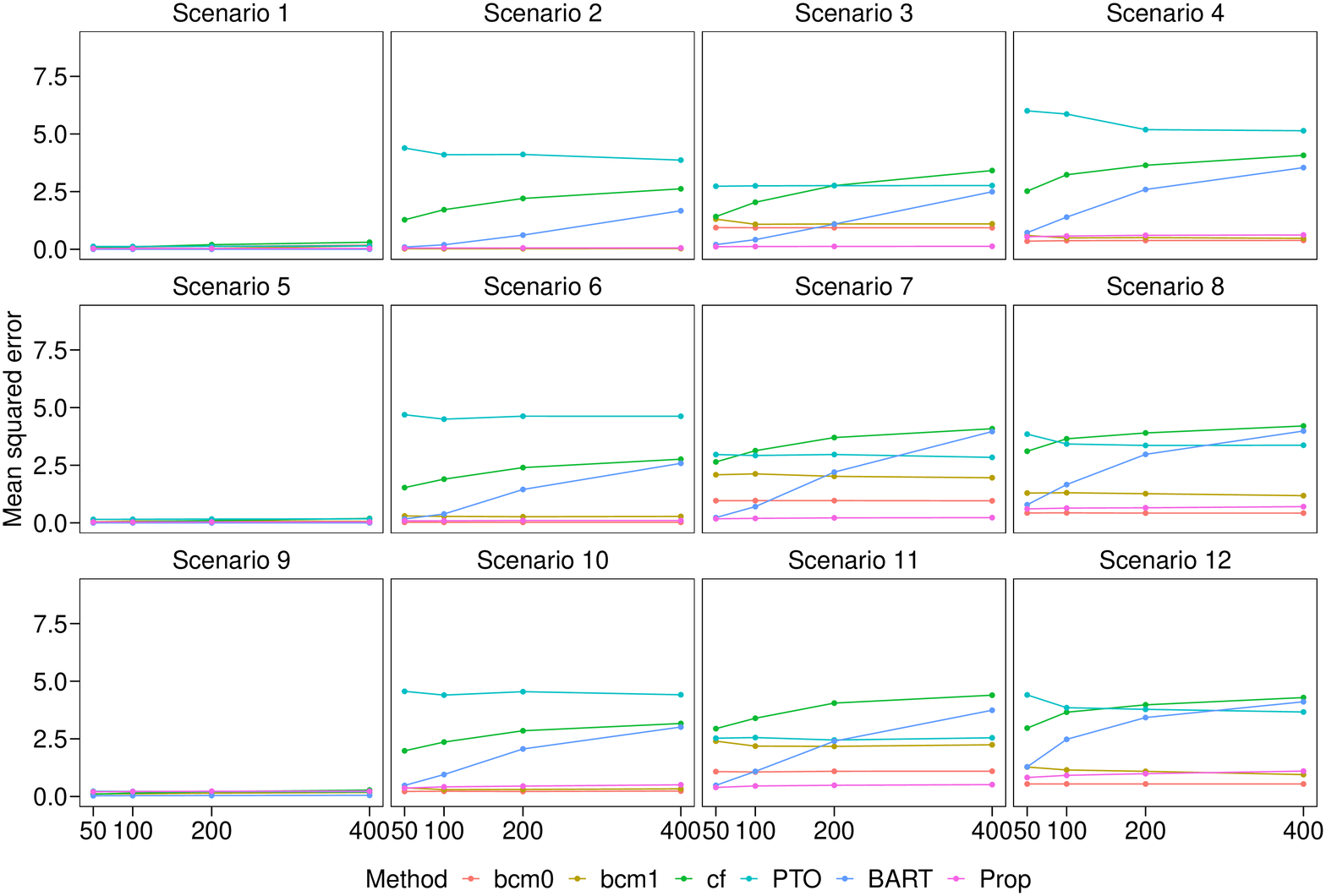}
 \caption{Results of simulated observational study datasets of twelve scenarios. For details of the generation model, see Table \ref{HTE.model}. The x-axis is the number of covariates in the simulation dataset for the corresponding scenario. The y-axis is the value of the mean squared errors (MSEs). These line charts show the median value of MSEs between the true and estimated the HTE with six different methods, including bcm0 = bagged causal MARS, bcm1 = bagged causal MARS, cf = causal forest, PTO = pollinated transformed outcome forest, BART = Bayesian additive regression tree, Prop = proposed method.}
 \label{obv_change}
\end{figure}

The results of the simulated observational study datasets of the twelve scenarios are shown in Fig.\ref{obv_change}. Here, we show the results of bagged causal MARS with and without adjustment of the propensity score. However, adjusting the bagged causal MARS’ propensity score for observational study data does not improve its prediction accuracy. Thus, regarding the observational data simulation, we discuss only bagged causal MARS without propensity adjustment. The trend in the prediction accuracy of each method in simulation of observational study data was similar to that of the RCT data. The proposed method outperforms the previous methods in scenarios 3, 7, and 11. In scenarios 2,6, and 10, the prediction accuracy of the proposed method is similar to that of the bagged causal MARS in scenarios 2 and 6 but somewhat inferior to that of the bagged causal MARS in scenario 10. For scenarios 4, 8, and 12, the prediction accuracy of the proposed method was slightly less than that of the bagged causal MARS. Scenarios 10, 11, and 12 also reflect that complicated non-linear covariates affect the prediction accuracy of the proposed method. In the simulation of observational studies, the prediction accuracy of the proposed method still tends to be higher than that of the causal forest, BART, and PTO forest and remains stable as the number of covariates increases.

Summarizing the simulation for both the RCT and observational datasets, the main strength of the proposed method is its high prediction accuracy for HTEs with piecewise constant relationships among the covariates. Furthermore, the prediction accuracy of the proposed method remained stable as the number of covariates increased. However, the complicated non-linear covariate impact may interfere with the prediction accuracy of the proposed method. Compared to previous methods, the proposed method outperformed the causal forest, BART, and PTO forests in all the scenarios. They outperformed bagged causal MARS when the HTE has piecewise constant relationships among covariates. Otherwise, the bagged causal MARS shows high prediction accuracy when the HTE has linear and quadratic structures. Although the proposed model cannot completely cover the HTE structure in these conditions, the prediction accuracy is not significantly inferior to that of bagged causal MARS.

\section{Real Data application}\label{sec5}

The proposed method was applied to the dataset from AIDS Clinical Trials Group Protocol 175 (ACTG 175) \citep{Hammer1996}. ACTG 175 is a randomized clinical trial with four arms, zidovudine monotherapy, zidovudine plus didanosine, zidovudine plus zalcitabine, and didanosine monotherapy in adults infected with human immunodeficiency virus type 1 (HIV-1) whose CD4 cell counts were from 200 to 500 per cubic milliliter. Here, the CD4 count is a significant indicator for quantifying the risk of HIV-1 infection, and a low CD4 count indicates a high risk of HIV-1 infection. The ACTG 175 dataset is available in the R package \texttt{speff2trial} \citep{speff2trial}, and 1762 subjects whose baseline CD4 cell counts were between 200 and 500 per cubic milliliter were selected for analysis.

This study focused on comparing the treatment effects of zidovudine monotherapy and combination therapy with zidovudine plus didanosine. The 419 subjects who received zidovudine monotherapy were included in the control group, and 436 subjects who received combination therapy with zidovudine plus didanosine were included in the treatment group. The outcome was set as the relative difference between the 20$\pm$5 weeks CD4 cell count and the baseline CD4 cell count, and it is denoted as (20$\pm$5 weeks CD4 cell counts - baseline CD4 cell counts)/baseline CD4 cell counts. As in Tsiatis et al. (2008), 12 covariates were selected \citep{Tsiatis2008}. These consist of five continuous variables: CD4 cell count (cd40; cell/mm$^3$) at baseline, CD8 cell count (cd80; cell/mm$^3$) at baseline, age (years), weight (wtkg; kg), Karnofsky score (karnof; on a scale of 0-100), and seven binary variables hemophilia (hemo; 0=no, 1=yes), homosexual activity (homo; 0=no, 1=yes), race (0=white, 1=non-white), gender (0=female, 1=male), history of intravenous drug use (drugs; 0=no, 1=yes), history of antiretroviral therapy (str2; 0=naive, 1=experienced) and symptomatic indicator (symptom; 0=asymptomatic, 1=symptomatic). 

\begin{figure}[t]
 \centering
 \includegraphics[width = \linewidth]{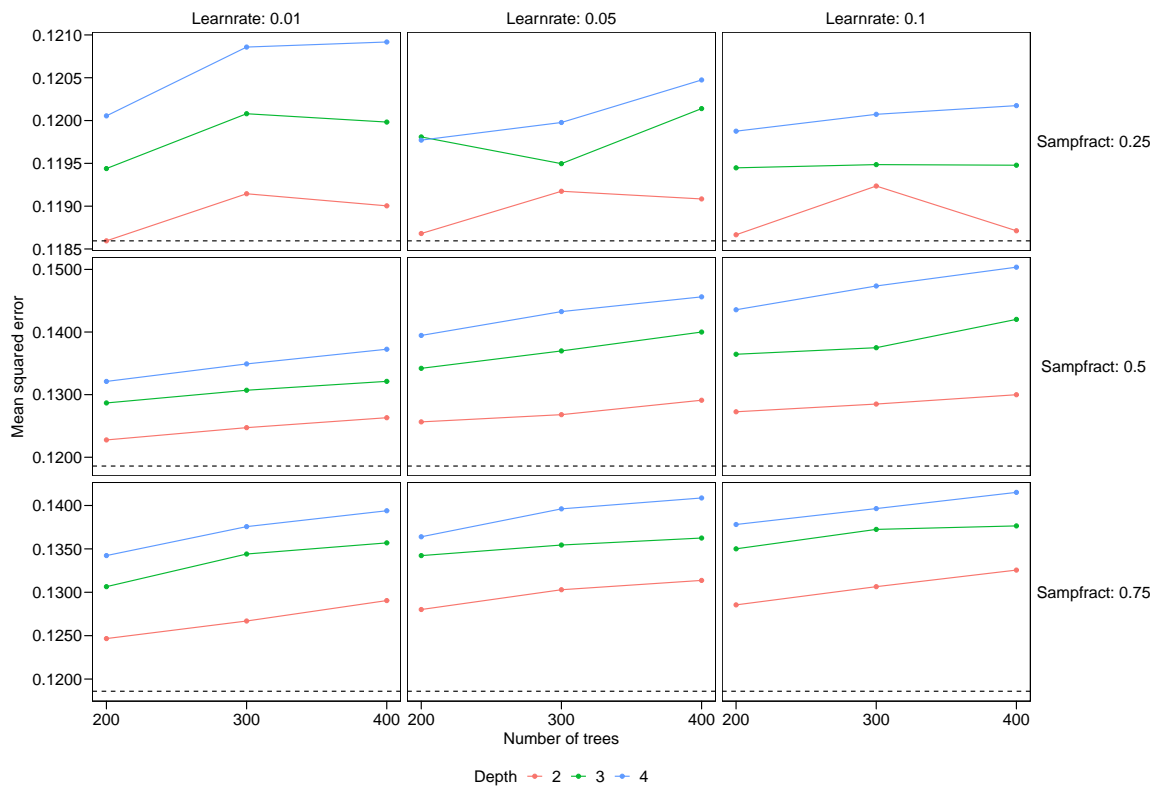}
 \caption{The mean squared error of the proposed method’s different hyperparameter combinations. The vertical axis of each chart is the mean squared error, and the horizontal axis of each chart is the number of tree-based learners in boosting. The charts for each column and row show the corresponding result of the shrinkage rate and the training sample fraction for each tree-based learner, respectively. In addition, the different colors indicate the mean depth of tree-based learners, and the black dashed lines indicate the minimum mean squared error among all the combination.}
 \label{cv_res}
\end{figure}

\paragraph{Parameter tuning for proposed method}
Before applying the proposed method, the hyperparameters must be determined. First, the candidate hyperparameters can be selected from the number of tree-based learners $M \in \{200, 300, 400\}$; mean depth of tree-based learners $\bar{L} \in \{2, 3, 4\}$; training sample fraction for each tree-based learner $\eta \in \{0.25, 0.50, 0.75\}$; and shrinkage rate $v \in \{0.01, 0.05, 0.10\}$. Then, 10-fold cross validation is repeated 30 times for parameter tuning, and the optimal hyperparameter combination that minimizes the mean squared error of the proposed model $F(\bm{x},t)$ (Eq.\ref{propmodel}) as
\begin{align}
MSE = \frac{1}{10}\sum_{b=1}^{10}\sum_{i=1}^N(y^{(b)}_i - \hat{F}^{(b)}(\bm{x}_i,t_i))^2
\end{align}
where $y^{(b)}_i$ and $\hat{F}^{(b)}(\bm{x}_i,t_i)$ are the true, and estimated outcomes of individual $i$ in $b$ cross-validation, respectively and $N$ is the sample size of the data. The mean squared errors for different combinations of hyperparameters are shown in Fig.\ref{cv_res}. Therefore, the optimized hyperparameter combination was determined as the number of tree-based learners $M = 200$, mean depth of tree-based learners $\bar{L} = 2$, training sample fraction for each tree-based learner $\eta = 0.25$, and shrinkage rate $v = 0.01$.
In addition, the ACTG 175 is RCT data, thus the propensity score for each individual can be simply determined as $\hat{\pi}(\bm{x}_i) = 0.5, i = 1, 2, \cdots, N$.

\begin{figure}[t]
 \centering
 \includegraphics[width = \linewidth]{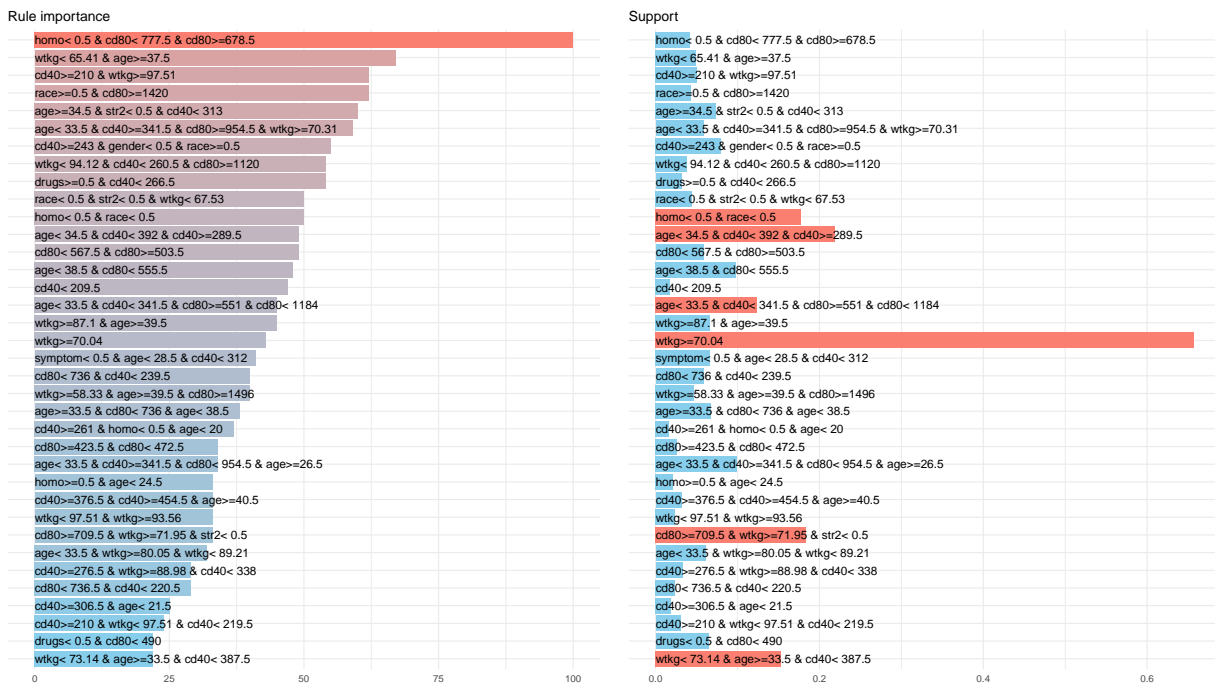}
 \caption{The bar charts reflect the importance of rules and their corresponding supports. The bar chart on the left, with its horizontal axis representing the importance of the rule, shows the rules with importance > the mean value. The bar chart on the right with the horizontal axis, represents the support of the rules, and the red bar indicates the rules with supports >0.1.}
 \label{RS}
\end{figure}

\paragraph{Application of proposed method}
By applying the proposed method with tuned parameters to the ACTG 175 dataset, we obtain 98 rules and their corresponding HTEs (coefficients). Thus, an HTE can be interpreted using the heterogeneity defined by these rules. However, because a direct interpretation of the effects of all rules would lead to complex results, we focus on rules that are more relevant to the HTE and can represent general results, thus allowing for easier comprehension. Therefore, we calculated the importance of each rule using Eq.\ref{imp.rule} or Eq.\ref{imp.linear}, and their corresponding support as Eq.\ref{support}, and the result is shown in Fig.\ref{RS}.  
The higher the importance of a rule, the more it affects the HTE. Support for a rule indicates the percentage of subjects who follow it; if the support is too small, it may be challenging to represent the general results. Therefore, we focused on rules with importance > the mean and with support >0.1.

Finally, six rules were selected, as shown in Table \ref{rule.res}, and are described in detail. The first rule is “homo $<$ 0.5 \& race $<$ 0.5” and its positive coefficient indicates that whites who do not participate in homosexual activity benefit from the combination therapy. The second rule “age$<$34.5 \& cd40$<$392 \& cd40$\geq$289.5” and its positive coefficient indicates that the subjects aged <34.5 and with baseline CD4 cell counts between 289.5 and 392 benefit from the combination therapy. The third rule “age$<$33.5 \& cd40$<$341.5\& cd80$\geq$551 \& cd80$<$1184" has a negative coefficient, thus, subjects aged <33.5, with baseline CD4 cell counts <341.5 and baseline CD8 cell counts between 551 and 1184 benefit from the monotherapy. The fourth rule “wtkg$\geq$70.04” has a positive coefficient, thus, subjects weighing >70.04 kg benefit from the combination therapy. The fifth rule “cd80$\geq$709.5 \& wtkg$\geq$71.95\&str2$<$0.5” also has a positive coefficient, thus the subjects with baseline CD8 cell counts >709.5, weight >71.95 and have not received the antiretroviral therapy benefit from the combination therapy. The last rule “wtkg$<$73.14 \& age$\geq$33.5\&cd40$<$387.5” has a positive coefficient, the subjects weighing <73.14, aged >33.5, and with a baseline CD4 cell count <387.5 benefit from the combination therapy.
\begin{center}
\begin{table}[]%
\centering
\caption{Result of the proposed method for the ACTG 175 study.\label{rule.res}}%
\begin{tabular*}{\linewidth}{@{\extracolsep\fill}cccc@{\extracolsep\fill}}
\toprule
\textbf{Rule Importance} & \textbf{Coefficient}  & \textbf{Support}  & \textbf{Rules}   \\
\midrule
50   & 0.06  & 0.18 & homo $<$ 0.5 \& race $<$ 0.5      \\
49   & 0.06   & 0.21 & age$<$34.5 \& cd40$<$392 \& cd40$\geq$289.5    \\
45   & -0.07   & 0.12 & age$<$33.5 \& cd40$<$341.5\& cd80$\geq$551 \& cd80$<$1184     \\
43   & 0.04   & 0.65 & wtkg$\geq$70.04   \\
33   & 0.04   & 0.18 & cd80$\geq$709.5 \& wtkg$\geq$71.95\&str2$<$0.5    \\
22   & 0.03   & 0.15 & wtkg$<$73.14 \& age$\geq$33.5\&cd40$<$387.5   \\
\bottomrule
\end{tabular*}
\end{table}
\end{center}

Summarizing the results of our proposed method, five of these rules show positive coefficients, therefore we obtained five subgroups that could benefit more from combination therapy with zidovudine and didanosine for HIV. The most important subgroup is that of white people who do not participate in homosexual activities. Similar results have also been reported by \cite{Spanbauer2021}, indicating that white people who do not participate in homosexual activities benefit from combination therapy. In most subgroups, combination therapy was superior to zidovudine monotherapy, indicating that combination therapy is likely to be more effective than monotherapy. This is also related to the results of \cite{Hammer1996}, who indicate that the average treatment effect of combination therapy was more effective than monotherapy. Furthermore, our proposed approach also identified subgroups for which zidovudine monotherapy was more effective. 

\section{Conclusion}\label{sec6}
In this study, we proposed an ML method based on RuleFit for HTEs that maintains the advantages of the original RuleFit and is interpretable and exhibits non-inferior accuracy compared to previous methods. The proposed method can also be regarded as an extension of RuleFit for estimating HTEs, defined in the potential-outcome framework. The proposed method refers to the “pollination” procedure in PTO forest, which creates base functions based on the transformed outcome. It then aggregates the average treatment effect within the terminal node of each tree-based function to estimate HTEs \citep{Powers2018}. Thus, the proposed method creates rules based on the transformed outcome, prunes the rules, and calculates the average treatment effect within the rules using the group lasso. In this way, the rules unrelated to the HTEs are removed. Furthermore, the HTEs for subgroups defined by the rules can be used to interpret the results. Compared to previous tree-based ensemble methods such as causal forest and BART, the proposed method produced a model by linearly combining rules and several linear terms instead of aggregating many tree models. 

Several simulation studies were conducted to compare the prediction accuracy of our method with that of previous studies. The proposed method outperformed the tree-based ensemble methods causal forest, BART, and PTO forest and exhibited non-inferior prediction accuracy to bagged causal MARS in RCT and observational data. In particular, the proposed method performed well for data with step-function relationships between the HTEs and covariates. In addition, the prediction accuracy of the proposed method remained stable with an increasing number of covariates compared with the tree-based ensemble methods. However, the proposed method is disadvantaged to the non-linear structure covariate effect or the relationship between the HTEs and the covariates compared to bagged causal MARS.

The proposed method was also applied to real data from an HIV study, ACTG 175. We also provided a process to evaluate the importance of rules to focus on additional rules related to HTEs. Therefore, in this application, we computed the importance of the rules and focused on those with importance greater than the average importance of the rule. In addition, to avoid irregular results, we selected rules with support more significant than 0.1. We obtained simple rules to interpret the clinical relevance between the HTEs and patient characteristics.

In summary, the proposed method demonstrates an interpretable model and prediction accuracy similar to previous methods. However, the interpretation of the proposed model has several limitations. Although the proposed method uses the group lasso to remove the rules unrelated to the HTEs, it can retain too many rules in the final model and complicate interpretation. In addition, we only focused on the HTEs in potential outcome frameworks and assumed continuous outcomes in this study. However, the HTE estimations for binary or survival outcome data are also reported \citep{Imai2013,Hu2021}. Therefore, future work is required to extend the proposed method to binary and survival outcomes.

\section*{Data Availability Statement}
The data used in Section 5 is available from R package \texttt{speff2trial} version 1.0.4. \\ https://cran.r-project.org/web/packages/speff2trial/

\subsection*{Conflict of interest}

No conflicts of interest.

\bibliographystyle{unsrtnat}
\bibliography{references}  






\end{document}